\documentclass[pdflatex,sn-mathphys-num]{sn-jnl}

\usepackage{graphicx,lscape, geometry}%
\usepackage{multirow}%
\usepackage{amsmath,amssymb,amsfonts}%
\usepackage{amsthm}%
\usepackage{mathrsfs}%
\usepackage[title]{appendix}%
\usepackage{xcolor}%
\usepackage{textcomp}%
\usepackage{manyfoot}%
\usepackage{booktabs}%
\usepackage{algorithm}%
\usepackage{algorithmicx}%
\usepackage{algpseudocode}%
\usepackage{listings}%
\usepackage[normalem]{ulem}%
\usepackage{hyperref}
\usepackage{pdfpages}
\usepackage{pax}

\theoremstyle{thmstyleone}%
\theoremstyle{thmstyletwo}%

\theoremstyle{thmstylethree}%

\begin{document}

\title[Article Title]{\textbf{One-milligram torsional pendulum toward experiments at the quantum-gravity interface} }


\author[1]{\fnm{Sofia} \sur{Agafonova}}
\author[1]{\fnm{Pere} \sur{Rossello}}
\author[1]{\fnm{Manuel} \sur{Mekonnen}}
\author*[1]{\fnm{Onur} \sur{Hosten}}\email{onur.hosten@ist.ac.at}

\affil[1]{\orgname{Institute of Science and Technology Austria}, \city{Klosterneuburg}, \country{Austria}}

\abstract{Probing the possibility of entanglement generation through gravity offers a path to tackle the question of whether gravitational fields possess a quantum mechanical nature. A potential realization necessitates systems with low-frequency dynamics at an optimal mass scale, for which the microgram-to-milligram range is a strong contender. Here, after refining a figure-of-merit for the problem, we present a 1-milligram torsional pendulum operating at 18 Hz. We demonstrate laser cooling its motion from room temperature to 240~microkelvins, surpassing by over 20-fold the coldest motions attained for oscillators ranging from micrograms to kilograms. We quantify and contrast the utility of the current approach with other platforms. The achieved performance and large improvement potential highlight milligram-scale torsional pendulums as a powerful platform for precision measurements relevant to future studies at the quantum-gravity interface.}





\maketitle

\subsection*{Introduction}

All fundamental forces apart from gravity have been demonstrated to adhere to quantum-mechanical laws. Aligning gravity with quantum mechanics has proved formidable to the point that it has sparked numerous proposals for alternative theories in which gravity is inherently non-quantum \cite{Grossardt2022,oppenheim2023}. Testing whether gravitational interactions between objects can generate quantum entanglement offers a first glimpse at the putative quantum nature of gravitational fields \cite{Bose2017,Marletto2017,kafri2013}. Currently, achieving a regime in which gravity dictates the interaction between two quantum-mechanically-behaving objects is an actively pursued challenge \cite{bose2023}. In this regime, quantum uncertainties in the position of a mechanical object should lead to quantum uncertainties in the gravitational fields sourced \cite{DeWitt2011}. 

While femto-to-microgram-scale oscillators have allowed observation of quantum mechanics in macroscopic objects \cite{9_Aspelmeyer2014}, and gram-to-ton-scale oscillators have enabled exploration of gravitational wave detection \cite{Abbot2016,PhysRevLett.99.160801}; the microgram-to-milligram scale \cite{Westphal2021,fuchs2024,Catano-Lopez2020,PhysRevLett.122.071101,PhysRevLett.131.043603,PhysRevA.106.013514,PhysRevA.101.011802,9329144,agatsuma2014,Michimura2020,Shin:25, Pluchar:25} is arguably the ideal range to probe the quantum-gravity interface. The reasons are simple: Quantum control is challenging with larger objects, while significant gravitational interaction is difficult with smaller objects. The feasibility of such experiments is not unwarranted with today's technology, but the challenges are demanding \cite{aspelmeyer2022}. Any progress in overcoming these challenges will undoubtedly drive progress also in studies of optomechanical entanglement \cite{PhysRevLett.98.030405,muller2008}, quantum decoherence \cite{PhysRevA.59.3204,PhysRevLett.91.130401}, wave function collapse \cite{PhysRevLett.114.050403}, the nature of dark matter \cite{Carney_2021}, and potential modifications to Newtonian gravity \cite{qvarfort2022,timberlake2021}.

Gravitational entanglement tests can be carried out with pairs of particles, either each in a discrete superposition state with superposition separation $\Delta x$ \cite{Bose2017,Marletto2017}, or with each in a coherent wavepacket of spatial spread $\Delta x$ \cite{balushi2018,Krisnanda2020,weiss2021,Miao2020,miki2024} --- equivalent to a continuum of superpositions. The entanglement rate of the particles is invariably given by $\Gamma_\text{ent}=\Delta x^2 |\nabla F|/\hbar$, where $|\nabla F|$ is the inter-mass force gradient \cite{bengyat2023,aspelmeyer2022,weiss2021,Krisnanda2020} --- e.g., for two spheres with mass $m$ each, $|\nabla F|=2Gm^2/d^3$, where $G$ is the gravitational constant, $d$ is particle separation, and $\hbar$ is the reduced-Planck constant. To extend the discussion to real-world systems in presence of thermal noise, we will adopt the concept of quantum coherence length \cite{huyet2001}, physically corresponding to the spatial distance $\xi$ over which an object can show quantum interference effects. In this context, a thermalized oscillator of angular frequency $\omega_0$ has a coherence length of $\xi=x_{\text{zp}}/\sqrt{(2n+1)}$, a value decreasing from its zero-point value $x_{\text{zp}}\equiv\sqrt{\hbar/2m\omega_0}$ with increasing mean excitation number $n$ (see Supplementary Section 2.1). Note that, in the zero-temperature limit, $\xi$ by definition reduces to $\Delta x$ defined in the context of entanglement rate. 

\begin{figure*}[t!]
    \centering
    \includegraphics[width=1.0\textwidth]{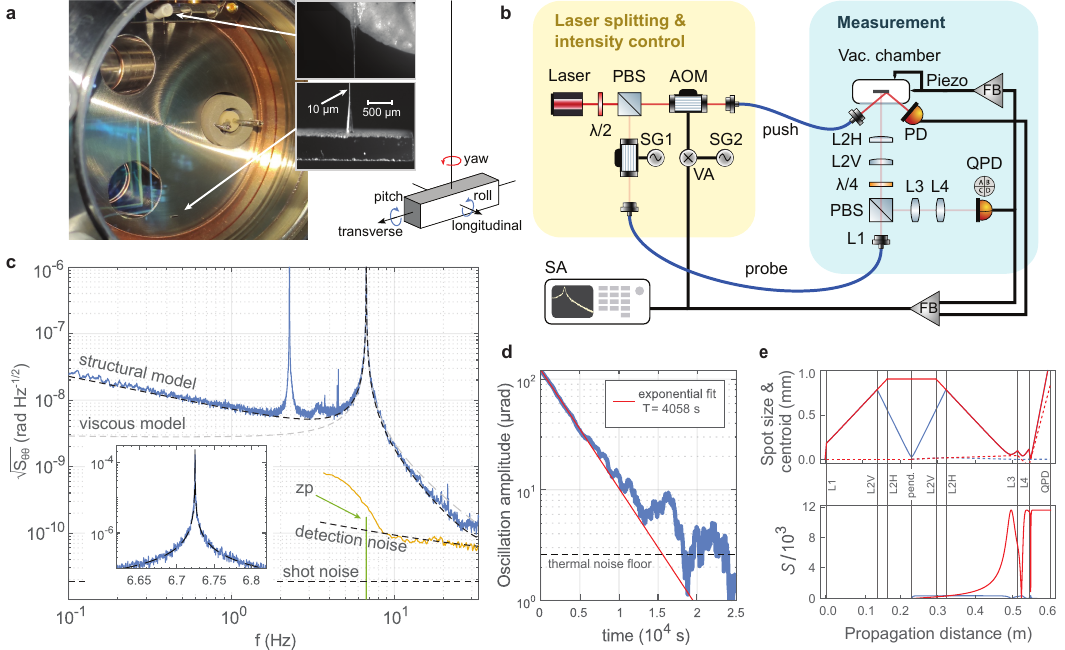}
    \caption{\textbf{Experimental setup.} 
    \textbf{a} The in-vacuum pendulum. Insets: Tapered suspension fiber ends and the pendulum mirror bar. \textbf{b} The optical setup. PBS: polarizing beam splitter, $\lambda/2$: half-wave plate, $\lambda/4$: quarter-wave plate, AOM: acousto-optic modulator, SG: signal generator, VA: variable attenuator, SA: spectrum analyzer, FB: feedback circuit, PD: photodetector, QPD: quadrant PD, L2H(V): horizontal (vertical) cylindrical lens with 200 mm (60 mm) focal length, L1(3, 4): spherical lens with 4.5 mm (10 mm, 4.0 mm) focal length. \textbf{c} Steady-state yaw noise spectrum (blue) compared to thermal noise models (dashed lines) with structural or viscous damping. Inset: close-up of the resonance peak, orange data: detection noise spectrum compared to theoretical laser shot-noise \cite{PhysRevX.13.011018}, ``zp'' curve: theoretical spectrum for quantum zero-point fluctuations associated with the yaw motion for reference. The parasitic 2.27-Hz peak is due to the residual swing motion sensitivity in the yaw channel. The ``detection noise'' model incorporates a variable ``white + 1/f'' noise to fit the data above 8 Hz. \textbf{d} Ring-down measurement of the 6.72-Hz yaw oscillations. \textbf{e} Probe beam profile across the lens system for horizontal (red) and vertical (blue) directions. Top; solid lines: spot size evolution, dashed lines: centroid evolution for a 100-$\mu\text{rad}$ illustrative yaw (red) or pitch (blue) rotation of the pendulum. Bottom; tilt sensitivity parameter $\mathcal{S}$.}
    \label{fig:setup}
\end{figure*}

The conditions for achieving gravitational entanglement between two continuously monitored optomechanical oscillators have been studied in references \cite{Miao2020} and \cite{miki2024} using logarithmic negativity to identify the onset of quantum entanglement in presence of dissipation. In terms of the coherence length $\xi$, this condition takes the form 
\begin{equation}\label{eq:entanglement rate}
    \eta^2=\frac{\xi^2 |\nabla F|}{\hbar\gamma}\geq1,
\end{equation}
where $\gamma$ is the oscillators' total mechanical energy dissipation rate. Note that the figure-of-merit $\eta$ can alternatively be written as $\eta^2\equiv\Gamma_\text{ent}^{(0)}/2\Gamma_{\text{dec}}$, where $\Gamma_\text{ent}^{(0)}$ is the entanglement rate when the oscillators are in their ground states, and $\Gamma_{\text{dec}}$ is the decoherence rate---see Supplementary Section 2.3.1.

For weakly interacting objects subjected to noise, Eq.~\ref{eq:entanglement rate} is satisfied in a regime where the purity $\mathcal{P}$ of the quantum controlled system approaches unity \cite{Krisnanda2020,miki2024}. Note that, for a thermal state of an oscillator $\mathcal{P}=\xi^2/x_{\text{zp}}^2=1/(2n+1)$. While Eq.~\ref{eq:entanglement rate} is the requirement for entanglement, objects become entangled only when quantum control enables $\mathcal{P}\rightarrow 1$. Together, $\eta$ and $\mathcal{P}$ form an experimentally relevant pair of quantities for the utility of a given oscillator at the quantum-gravity interface, allowing a comparison of many existing platforms. It should be noted that, aside from direct entanglement generation, there exist tests at the quantum–gravity interface that do not require $\eta>1$ or $\mathcal{P}\rightarrow 1$, yet remain characterized by these parameters. For example, when probing for a necessary noise source that would prevent entanglement generation for non-quantum gravity \cite{angeli2025,oppenheim2023gravitationally,kryhin2025distinguishable,blencowe2013effective}, $\eta$ sets the strength of the noise. Similarly, if squeezing in the differential position of the interacting objects can be taken as a signature of the quantum nature of gravity \cite{datta2021,miki2023}, $\eta$ enters in the level of squeezing.

In this article, we present the first generation of an experiment aimed at the quantum-gravity interface, where we develop and manipulate a 1-milligram suspended torsional pendulum. We demonstrate laser cooling of the pendulum motion through radiation-pressure feedback, suppressing the thermal motion from room temperature to an effective temperature of $240~\mu\text{K}$. Cooling is achieved by first optically trapping the torsional motion to shift its frequency from 6.72~Hz to 18.0~Hz, and then by critically damping the motion. With the developed system, we benchmark a quantum-gravity figure-of-merit $\eta=4\times10^{-9}$ and a purity $\mathcal{P}=1.4\times 10^{-6}$ that we will put into context below. The achieved cooling levels significantly surpass the lowest reported motional temperatures in the microgram-to-kilogram mass window --- 6~mK for the gram scale \cite{PhysRevLett.99.160801} and 15~mK for the milligram scale \cite{PhysRevA.94.033822} --- and the achieved torque sensitivity of $1.2\times10^{-18}$ N~m~Hz\textsuperscript{-1/2} constitute a record value for the milligram scale -- a factor of 10 beyond reference \cite{PhysRevA.101.011802}. For a detailed discussion of a concrete example to realize gravitational entanglement between two pendulums with systems similar to the one pursued here, see \cite{miki2024}. There, the feasibility of steady-state entanglement generation is analyzed, highlighting the importance of measurement-based state conditioning and feedback cooling in the presence of relevant decoherence effects.

\subsection*{Setup}

The pendulum used in this study is shown in Figure~\ref{fig:setup}(a) inside its vacuum housing maintained at 5$\times$10\textsuperscript{-9}-mbar. The pendulum consists of a $2\times0.5\times0.5$~mm\textsuperscript{3} silver-coated fused silica mirror bar, and a 10-$\mu\text{m}$ diameter fused silica suspension fiber of 5-cm length. It exhibits a torsional (yaw) resonance at $f_0\equiv\omega_0/2\pi=6.72$~Hz in addition to longitudinal and transverse swing modes at 2.27~Hz, a roll mode at 44.48~Hz, and a pitch mode at 125.95~Hz (Fig.~\ref{fig:setup}(a), overlay). The piece used for the pendulum bar was cut from a larger optical mirror, and the suspension fiber was manufactured in-house from a standard 125-$\mu\text{m}$-diameter optical fiber (S630-HP) by wet etching. The etching process was designed to leave 500~$\mu\text{m}$ long tapers at both ends (Fig.~\ref{fig:setup}(a), inset) to allow adhesive-aided attachment without introducing mechanical losses. In this way, the torsional energy is mainly stored in the high-compliance thin section of the fiber, preventing dissipation at the low-compliance 125-$\mu\text{m}$ diameter adhesive-bonded (Norland 61) surfaces. This design achieves a quality factor of $Q_0=\tau \omega_{0}/2=8.6\times10^{4}$ for the yaw mode, as measured by a ring-down of its oscillation amplitude with a time constant of $\tau=4058$~s (Fig.~\ref{fig:setup}(d)). The ring-down thermally equilibrates around $(k_\text{B} T_0/(I\omega_0^2))^{1/2}=2.6~\mu\text{rad}$ fluctuations dictated by the equipartition theorem. Here; moment of inertia $I=3.3\times10^{-13}$ kg~m\textsuperscript{2}, ambient temperature $T_0=295~\text{K}$, Boltzmann constant $k_\text{B}$. The measured $Q_0$ and the corresponding mechanical energy dissipation rate of $\gamma_0=\omega_0/Q_0=2\pi\times 79~\mu\text{Hz}$ indicate excellent performance even compared to torsional modes of monolithic pendulums manufactured with laser-welded suspensions \cite{Catano-Lopez2020} --- with room only a factor of 2-3 to reach material limited dissipation based on the the utilized fiber's surface-to-volume ratio \cite{gretarsson1999}. To appreciate the potential of the torsional mode of this system for quantum experiments, it is crucial to understand that $Q_0$ is not the final quality factor, but that it can be increased many orders of magnitude by increasing the oscillation frequency with an optical spring, while remaining at the same material quality factor \cite{Michimura2020}.

The near-ideal thermally-limited noise spectrum (Fig.~\ref{fig:setup}(c)) reveals that dissipation in the yaw motion arises purely from structural damping, with no observable viscous contribution from background gas. This spectrum arises from a thermal (th) torque ($\tau$) noise of power spectral density (PSD) $S_{\tau\tau}^{\text{th}}=4k_\text{B} T_0 I \gamma(\omega)$ with frequency-dependent structural dissipation rate $\gamma(\omega)=\gamma_0\frac{\omega_0}{\omega}$ \cite{Michimura2020}. The torque noise manifests itself as an angular ($\theta$) PSD $S_{\theta\theta}^{\text{th}}=|\chi|^2 S_{\tau\tau}^{\text{th}}$ through the mechanical susceptibility $\chi(\omega)=(\omega_0^2-\omega^2+i\omega\gamma(\omega))^{-1}/I$, showing excellent agreement with the measurements. The agreement in the peak region (Fig.~\ref{fig:setup}(c), inset) indicates a resonance frequency instability below $100~\mu\text{Hz}$. The minimal observed torque noise of $1.2\times10^{-18}$ N~m~Hz\textsuperscript{-1/2} makes the system excel as a competitive sensor (Supplementary Section 1.6).

The characterization of the motion is enabled by an optical lever capable of resolving the motion at the level of quantum zero-point fluctuations $S_{\theta\theta}^{\text{zp}}=S_{\theta\theta}^{\text{th}}/(2n_{\text{th}}+1)$ that would be associated with the yaw oscillations (Fig.~\ref{fig:setup}(c)). Here, $n_{\text{th}}\approx k_\text{B} T_0/\hbar\omega_0=9.2\times10^{11}$ is the mean number of thermal excitation quanta. The calibration of the optical lever signal was done by replacing the pendulum with a rigid rotatable mirror. Based on this calibration, the measured yaw angle noise agrees with the expected thermal noise levels to within 10\%. The detection noise spectrum shown in Figure~\ref{fig:setup}(c) is the base noise of the optical lever obtained in this rigid mirror configuration. The excess over shot noise level is due to the lever's own beam pointing fluctuations; see Supplementary Section 1.5 for additional characterizations of the detection noise.

The optical setup around the pendulum is shown in Figure~\ref{fig:setup}(b). The output of a 780-nm DFB laser is split into two paths: one to probe the pendulum's motion (10~$\mu\text{W}$) and the other to manipulate it through radiation pressure (0-4~mW). Accousto-optic modulators (AOM) enable intensity control on each path. The Gaussian probe beam is incident at the center of the pendulum, and circulates back to a quadrant photodiode (QPD), which reads the horizontal and vertical positions of the beam to yield the pendulum motion. 

An arrangement of cylindrical and spherical lenses maximizes the sensitivity to yaw motion (horizontal beam tilt), while minimizing the sensitivity to any type of mechanical mode that leads to a pitching-like motion (vertical beam tilt). The beam shaping principle (Supplementary Section 1.1) is based on two considerations: 1) independently of beam divergence, the fundamental tilt sensitivity is linearly proportional to the beam spot size $\text{w}_\text{p}$ at the pendulum, and 2) the utilized tilt sensitivity parameter $\mathcal{S}=\sqrt{\frac{8}{\pi}}\frac{d}{d\theta}\frac{\delta(\theta)}{\text{w}}~~(\text{rad}^{-1})$ can be maximized to $\mathcal{S}_\text{max}=\tfrac{\sqrt{32\pi}}{\lambda}\text{w}_\text{p}$ for any desired beam size at the detection point, extracting the tilt information optimally. In the definition of the sensitivity parameter, $\text{w}$ and $\delta(\theta)$ are the spot size and the tilt-induced beam displacement at the detection point; and $\lambda$ is the wavelength---see Supplementary Section 1.1 for additional discussions on $\mathcal{S}$. The implemented probe beam profile and the resulting $\mathcal{S}$ as a function of detection distance are illustrated in Figure~\ref{fig:setup}(e). $\mathcal{S}$ is maximized in the horizontal channel and minimized in the vertical one, both for a large spot size at the detection point, avoiding photodiode gaps. 

The QPD output is fed into a home-built analog feedback circuit to control the push beam power (Supplementary Section 1.2) incident on the corner of the pendulum to implement effective equations of motion for the yaw mode (Fig.~\ref{fig:setup}(b)). Although the detection is optimized for the yaw motion, smaller signals from all other modes of the pendulum are also available on the QPD output. We use these to dampen those other modes via feedback by pushing on the vacuum chamber with piezo-actuators (Supplementary Section 1.4).

\begin{figure*}[t!]
    \centering
    \includegraphics[width=1.0\textwidth]{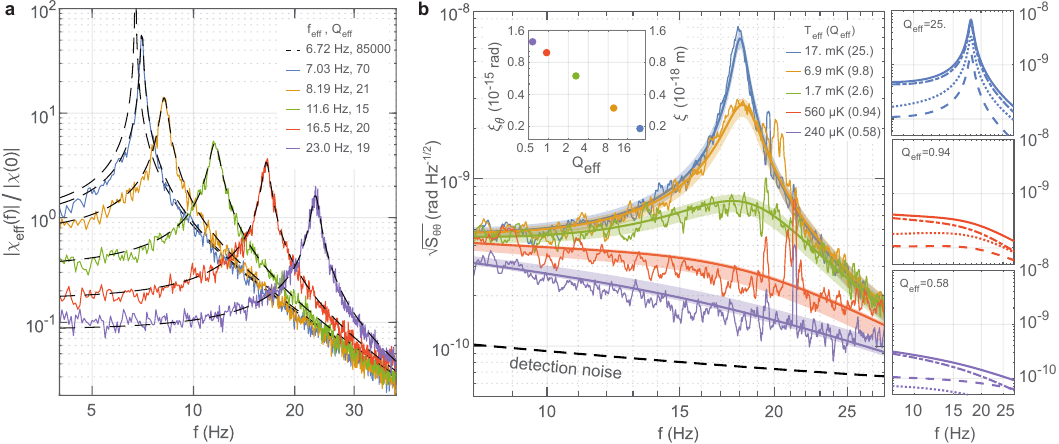}
    \caption{\textbf{Feedback control of the motion.} \textbf{a} Experimental effective susceptibilities as the resonance frequency of the pendulum is shifted under feedback control. Normalization is to the DC value of the feedback-free susceptibility. \textbf{b} Noise spectral densities for different feedback damping strengths for the pendulum shifted to 18 Hz. Data points: mean spectral density within each 2.4-Hz frequency bin, error bars: $\pm1$ standard deviation within each bin, solid lines: theoretical noise models with excess vibration noise as a fit parameter, error bands: variation in the model curves due to the full-range of lab vibration noise level changes, detection noise: same as Fig.~\ref{fig:setup}. Inset: coherence angles $\xi_\theta$ and pendulum-tip coherence lengths $\xi$. Side panels show the noise breakdown for representative damping strengths. Solid lines: total noise, dash-dotted lines: theoretical thermal noise given the effective susceptibilities, dashed lines: theoretical imprinted measurement noise, dotted lines: extracted white-torque-noise equivalent vibration noise.}
    \label{fig:results}
\end{figure*}

\subsection*{\label{sec:OptMechControl}Optomechanical control}
The feedback-based radiation forces are designed to implement the effective susceptibility function 
\begin{equation} \label{eq:eff_susceptibility}
    \chi_{\text{eff}}(\omega)=(\omega_{\text{eff}}^2-\omega^2+i\omega\gamma_{\text{eff}})^{-1}/I
\end{equation}
of a torsional harmonic oscillator with tunable frequency and damping. The analog feedback circuit implementing the effective susceptibility contains two key parameters: gain of the feedback loop $g$ and the lead-filter corner frequency $\omega_{\text{lead}}$. Physically, while the gain controls the strength of the applied restoring torque, the lead filter introduces the time derivative of the measured yaw angle, implementing velocity damping (Supplementary Section 1.3). The parameters dictate the effective resonance frequency $\omega_{\text{eff}}=\omega_0\sqrt{1+g}$ and the effective damping rate $\gamma_{\text{eff}}=\frac{\omega_0^2 g}{\omega_{\text{lead}}}+\gamma(\omega)$. Here, the intrinsic damping $\gamma(\omega)$ is negligible relative to the feedback damping term, rendering the effective damping independent of frequency (purely viscous) with an effective quality factor of $Q_{\text{eff}}=\frac{\omega_{\text{eff}}}{\gamma_{\text{eff}}}$.

Figure~\ref{fig:results}(a) illustrates our ability to shift the torsional oscillation frequency through feedback-based optical forces, realizing an optical torsion spring. To experimentally characterize the effective susceptibilities, we optically apply a white-noise torque drive to the pendulum that is about 100 times the thermal torque noise. This creates a large motion, enabling an easy measurement of the angular noise spectrum to calculate $|\chi_\text{eff}|=(S_{\theta\theta}/S_{\tau\tau})^{1/2}$. During these measurements, we additionally induce a level of feedback-damping to stay below the maximal restoring torque that can be supplied by the finite power of the push beam. For each configuration, the effective frequencies and quality factors are extracted by fitting Eq.~\ref{eq:eff_susceptibility} to the data, obtaining excellent agreement with the intended effective susceptibilities.

Unlike the restoring torsion provided by the suspension fiber that brings in thermal torque noise, the optically induced torques can practically be noise-free \cite{PhysRevA.105.043520}. An oscillator whose frequency is increased by an optical spring acquires a new set of apparent parameters that dictate its properties. Defining the apparent quantities for the case of `no induced feedback damping', the susceptibility describing the oscillator becomes $\chi_\text{app}(\omega)=(\omega_{\text{eff}}^2-\omega^2+i\omega\gamma(\omega))^{-1}/I$, where $\gamma(\omega)$ is the original structural damping function. The resonant damping rate decreases to $\gamma_{\text{app}}=\gamma(\omega_\text{eff})=\gamma_0\frac{\omega_0}{\omega_{\text{eff}}}$, and the resonant quality factor increases to
\begin{equation}
    Q_{\text{app}}=\frac{\omega_{\text{eff}}}{\gamma_{\text{app}}}=Q_0\left(\frac{\omega_{\text{eff}}}{\omega_0}\right)^2.
\end{equation}
Provided that the detection noise floor does not impede, a larger apparent quality factor allows cooling to lower temperatures. For behavior beyond this regime---where detection noise is imprinted onto the motion--- see \cite{PhysRevA.105.043520} and Supplementary Section 1.3. To understand the improved cooling capability, note that feedback cooling sets the value of $\gamma_{\text{eff}}$ in Eq. \ref{eq:eff_susceptibility}, reducing the resonance peak height without adding noise. The largest meaningful value of  $\gamma_{\text{eff}}$ is achieved at critical damping $Q_{\text{eff}}=1/2$, after which point there is no oscillator. Therefore, one needs to start with a high quality factor to be able to achieve large cooling factors. 

To explore the limits of our ability to cool the yaw motion, we shift the yaw-mode resonance to 18~Hz, to the center of the excess-noise-free frequency band from 8 to 28~Hz. Below 8~Hz, spurious detection noise kicks in; above 28~Hz, excess noise of the 44-Hz roll motion leaks into the yaw-angle measurement. For the 18-Hz oscillator, $Q_\text{app}=6.1\times10^{5}$ and $\gamma_\text{app}=2\pi\times 29~\mu\text{Hz}$. After shifting the frequency, we implement five different feedback circuit configurations with effective damping rates corresponding to quality factors ranging from $Q_{\text{eff}}=25$ to $Q_{\text{eff}}=0.58$. We characterize the resulting effective susceptibilities using the same procedure described in the context of Figure~\ref{fig:results}(a). Then, for each feedback configuration, we let the system operate without any external driving torque. Figure~\ref{fig:results}(b) shows the yaw motion noise spectra obtained, showing the progressively colder oscillator states.

The resulting motion is governed by the conversion of three torque noise contributions to angular motion: structural thermal noise, detection noise that is imprinted on the motion through the feedback loop, and vibrations that affect the yaw motion (see Supplementary Section 1.3 for model details). Given the theoretical thermal noise, the characterized susceptibilities, and the detection noise, the only unknown is the vibration noise in the system --- known to be nonstationary. We take the vibration spectra to be approximately white noise (in torque) within the frequency band of interest and set its amplitude to be the fit parameter for the overall model. The resulting model curves are shown together with the spectral noise data in Figure~\ref{fig:results}(b). The noise breakdown is illustrated in the right panel of Figure~\ref{fig:results}(b) as well for exemplary cases: the system is driven dominantly by thermal noise, while the imprinted measurement noise and vibration noise are always subdominant.

The extent of the fluctuations in the yaw angle $\theta$ around zero can be expressed as an effective temperature $T_\text{eff}=I\omega_\text{eff}^2\langle \theta^2\rangle/k_\text{B}$ for this motion. In this context, the yaw angle fluctuations are determined by the integral of the angular PSD: $\langle \theta^2\rangle=\int_0^\infty S_{\theta\theta}(f)df$. The value of this integral is dominated by the contributions within the vicinity of the resonance peak. In fact, under the action of thermal noise alone, taking the PSD for the case of a purely frequency-shifted pendulum $S_{\theta\theta}^\text{app}=|\chi_\text{app}|^2S_{\tau\tau}^{\text{th}}$, one recovers to a good approximation the equipartition theorem $I\omega_\text{eff}^2\langle \theta^2\rangle_\text{app}\!\!=k_\text{B} T_0$ \cite{PhysRevA.105.043520,doi:10.1126/science.abh2634} --- already reaching to a 94\% accuracy even within a small integration band of $10\frac{\gamma_\text{app}}{2\pi}\equiv0.29~\text{mHz}$. Experimentally, the achieved effective temperatures can be determined by referencing the observed angular fluctuations within the accessible observation band to those of the ideal frequency-shifted oscillator:
\begin{equation} \label{eq:temperature}
    \frac{T_\text{eff}}{T_0}=\frac{\int_{f_1}^{f_2} S_{\theta\theta}(f)~df}{\int_{f_1}^{f_2} S_{\theta\theta}^{\text{app}}\!(f)~df}.
\end{equation}
In our case, $f_1=8~\text{Hz}$ and $f_2=28~\text{Hz}$. For each damping configuration in Figure~\ref{fig:results}(b), the temperatures extracted using Eq. \ref{eq:temperature} are indicated, reaching a minimum of $238\pm8~\mu\text{K}$ near critical damping. The uncertainty represents the 95\% confidence range of the model fit. This temperature is equivalent to a mean thermal excitation quanta of $n=2.8\times10^{5}$.

\subsection*{Discussion}

We now return to the figure-of-merit $\eta$ and the purity $\mathcal{P}$ in order to contextualize the results for the quantum-gravity interface and to get a comparison with experiments on various physical platforms. For this purpose, we first address the achievable gravitational force gradients $|\nabla F|$, then discuss the coherence lengths $\xi$ achieved in this work. But to begin with, we note that it is the total damping rate $\gamma_\text{eff}$ under feedback control that goes into the calculation of $\eta$. 

For two gravitationally interacting objects, the masses and geometry of the involved objects set a limit on how large the force gradient can become. For compact objects (e.g., spheres) the limit is reached when the objects come close enough to touch. For large aspect ratio objects (e.g., membranes, cantilevers, beams), however, the force gradient typically saturates when the separation approaches the larger of the object's dimensions due to the distributed nature of the resulting gravitational interaction. The saturation behavior is studied in Supplementary Section 2.3 for representative cases, and the results are utilized for the comparisons below. Even if object sizes themselves allow for separations below $50~\mu\text{m}$, such separations currently have no meaning for gravity experiments due to complications that arise in shielding Casimir and other parasitic electromagnetic interactions that would otherwise overshadow gravity. These shielding-related problems have been repeatedly encountered to date in gravity-related experiments, as well as in experiments with trapped ions in the form of electric field noise near conducting surfaces \cite{hite2021}. For example, although gold screens with 100-nm-level thicknesses can easily be manufactured for electromagnetic shielding, parasitic interactions with screens due to patch potentials to date have prevented studying gravity below $50~\mu\text{m}$ separations \cite{lee2020,blakemore2021}.

\begin{figure*}[t!]
    \centering
    \includegraphics[width=0.8\textwidth]{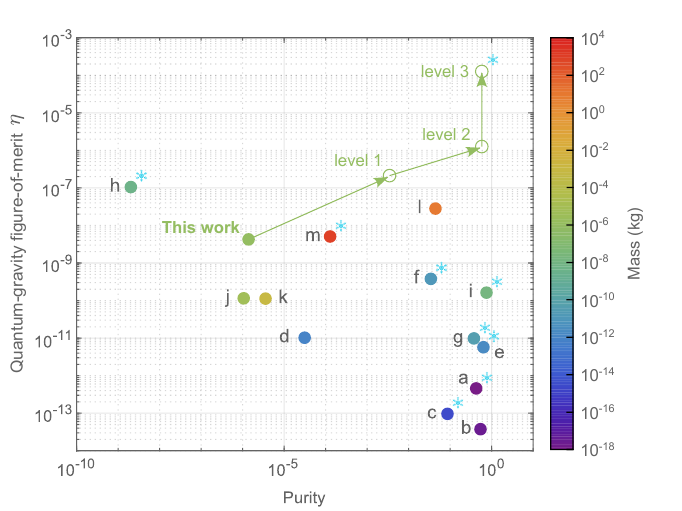}
    \caption{\textbf{Comparison to state-of-the-art mechanical quantum control experiments for utility at the quantum-gravity interface.} Colors encode oscillator masses. Experiments in a cryogenic environment are tagged with snow flakes. Experiments: \textbf{a} nanosphere \cite{rossi2024}, \textbf{b} nanosphere \cite{doi:10.1126/science.aba3993}, \textbf{c} nanobeam \cite{Wilson2015}, \textbf{d} microsphere \cite{monteiro2020}, \textbf{e} membrane \cite{Rossi2018}, \textbf{f} cantilever \cite{PhysRevLett.130.033601}, \textbf{g} membrane \cite{PhysRevA.92.061801}, \textbf{h} superconducting microsphere \cite{PhysRevLett.131.043603}, \textbf{i} acoustic resonator \cite{bild2023}, \textbf{j} pendulum \cite{PhysRevA.94.033822}, \textbf{k} pendulum \cite{PhysRevLett.99.160801}, \textbf{l} LIGO pendulums \cite{doi:10.1126/science.abh2634}, \textbf{m} bar resonator \cite{PhysRevLett.101.033601}. For `This work', projected future performances at different levels are also indicated; see main text.}
    \label{fig:comparison}
\end{figure*}
 

To obtain coherence length values, we first characterize coherence angles. Coherence angles can be found similarly to the coherence length of a linear oscillator, simply by replacing the linear zero-point fluctuations $x_{\text{th}}$ with its angular counterpart $\theta_{\text{th}}=\sqrt{\hbar/2 I \omega_\text{eff}}$ --- $I$: moment of inertia, $\theta_{\text{th}}=1.2\times10^{-12}~\text{rad}$ @ $\omega_\text{eff}=2\pi\times 18~\text{Hz}$. Compared to a standard damping mechanism through a thermal bath, a careful analysis for the case of feedback damping reveals an additional suppression in the coherence angle and purity (Supplementary Section 2.2):
\begin{equation} \label{eq:experimental coherence}
    \begin{split}
        \xi_\theta&= s~\theta_{\text{th}}/\sqrt{2n+1},\\
        \mathcal{P}&= s/(2n+1).
    \end{split}
\end{equation}
\noindent Here, $s=\bigl(1+Q_\text{eff}^{-2}~\Delta\theta_\text{fb}^2/\Delta\theta^2\bigr)^{-1/2}$ is the suppression factor, which has noticeable effects only when the oscillator becomes near-critically damped, with the effect increasing for overdamping. The ratio $\Delta\theta_\text{fb}/\Delta\theta$ is that of the feedback-imprinted-measurement-noise to the total-angular-noise, and can be extracted from the analysis in Figure~\ref{fig:results}(b). For our range of parameters, this suppression causes at most a 22\% reduction in coherence angles and purities from the simple picture that omits $s$. The coherence angles evaluated for our experimental configurations are shown in the inset of Figure~\ref{fig:results}(b). The implied coherence lengths $\xi$ for the tip of the pendulum bar at $d=1~\text{mm}$ from the rotation axis are also indicated. At the strongest damping, we obtain the largest coherence angle $\xi_\theta=1.2\times 10^{-15}~\text{rad}$, resulting in the coherence length $\xi=1.2\times 10^{-18}~\text{m}$ --- a thousandth of the size of an atomic nucleus.

Although $\xi$ might seem small, among physical systems that span more than 20 orders of magnitude in their masses, our intermediate-scale mass with its low frequency exhibits a remarkable performance with $\eta=4\times10^{-9}$ and $\mathcal{P}=1.4\times 10^{-6}$ when contrasted with other state-of-the-art mechanical quantum control experiments (Fig.~\ref{fig:comparison}): With respect to $\eta$, it is roughly on par with ton-scale bars and 10-kg-scale pendulums; and with respect to achieved purity $\mathcal{P}$, it is significantly ahead of micgrogram-scale superconducting spheres that currently suggest the largest $\eta$. We carefully lay out all the numbers and rationale that go into the comparisons in Supplementary Section 2.3.

Besides its current performance, the platform offers a large amount of improvement potential. First, note that feedback cooing only affects $\mathcal{P}$ (see Supplementary Section 2.3.1). On the other hand, decreasing noise by reducing dissipation directly affects $\eta$. Decreasing environmental temperature improves both quantities. We can foresee three possible levels of improvements that are indicated in Fig.~\ref{fig:comparison}, with all requirements (e.g., for ground state cooling) explained in detail in Supplementary Section 2.3.4. Here we provide a summary. Level~1: We can realize a 65-mHz torsional oscillator by switching to a 1-$\mu\text{m}$-diameter suspension fiber with a bare quality factor of $2\times 10^4$--- a readily demonstrated technology \cite{Catano-Lopez2020,tebbenjohanns2023}. Shifting the frequency to 18 Hz (as in this work) through feedback control will result in an apparent quality factor of $Q_\text{app}\sim1.5\times10^9$ and an apparent damping rate of $\gamma_\text{app}\sim2\pi\times 10~\text{nHz}$. Critical feedback damping will then allow another 3000-fold cooling, down to $n=100$. Level~2: It will likely be impractical to push the system further with optical lever detection. Utilizing optical cavities engineered for sensing torsional motion \cite{agafonova2024}, operating with a finesse of several thousands, can further allow cooling the system to its ground state. Level~3: Operating the system in a 30-mK dilution refrigerator can then boost $\eta$ by another 100-fold, while also reducing the required strengths of feedback cooling. Speculating beyond this performance would be uninformed at this point.

Deterministic entanglement between macroscopic objects has recently been demonstrated using RF circuits as mediators between 10-micron-scale superconducting drumheads \cite{kotler2021}. Our demonstrated platform advances the experimental capabilities required for exploring analogous correlations in gravitational systems. At the same time, the comparison of operating parameter ranges across existing approaches underscores how challenging this goal remains, suggesting that qualitatively new experimental strategies may be required. In the meantime, non-entanglement-based tests at the quantum–gravity interface may provide valuable guidance for future directions. Advancing macroscopic quantum optomechanics and precision torque sensing, our demonstrated platform offers a promising near-term path in this context.


\bmhead{Acknowledgements}We thank Gerard Higgins, Andrei Militaru, Nikolai Kiesel and Markus Aspelmeyer for useful discussions on the topic of the figure-of-merit. We thank Teodor Strömberg for helping with the additional characterizations of the optical lever noise. We thank Johannes Fink and Scott Waitukaitis for their helpful feedback on the manuscript. This work was supported by Institute of Science and Technology Austria, and the European Research Council under Grant No. 101087907 (ERC CoG QuHAMP). 

\bmhead{Author contributions}O.H. conceived the experiment and supervised the project. M.M. manufactured the pendulum. S.A. and P.R. built the experimental setup. S.A. and O.H. performed the reported experiments, analyzed the data, and wrote the article.

\bmhead{Competing interests}Authors declare no competing interests.


\bibliography{pendulum}

\includepdf[pages=-]{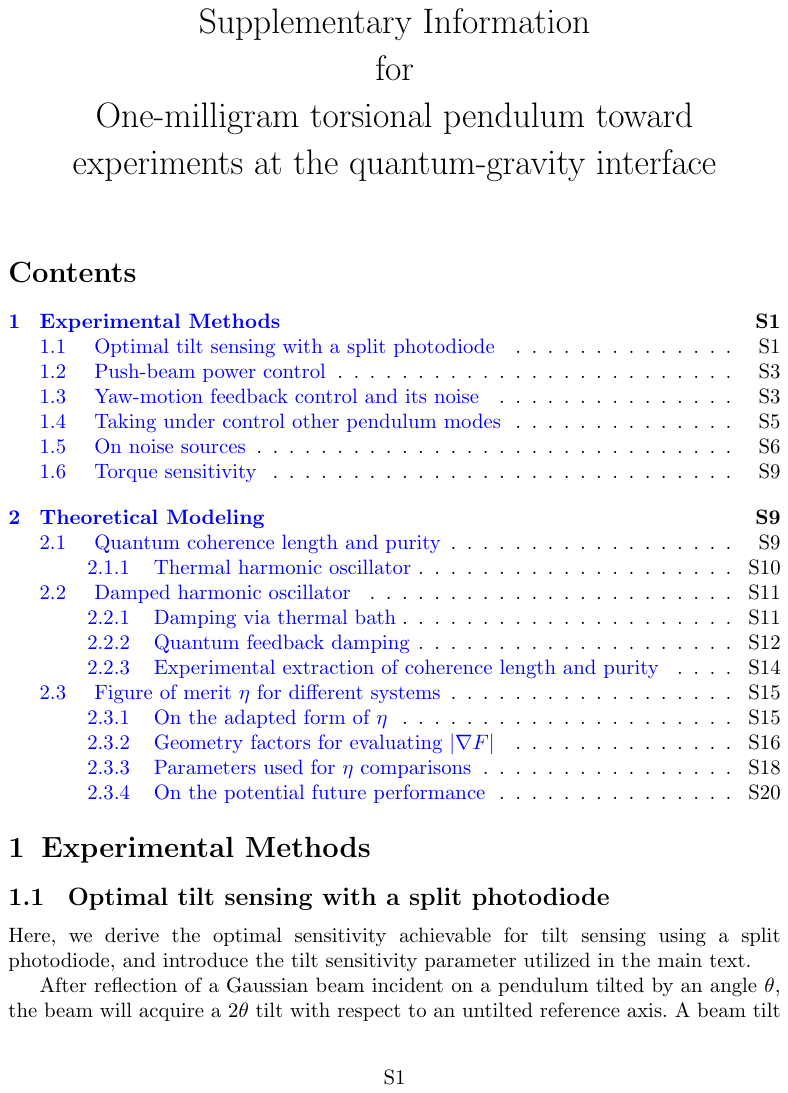}

\end{document}